\documentclass[a4paper]{article}

\usepackage[english]{babel}
\usepackage[utf8x]{inputenc}
\usepackage[T1]{fontenc}
\usepackage{multirow}
\usepackage[a4paper,top=1in,bottom=1in,left=1in,right=1in,marginparwidth=1in]{geometry}

\usepackage{amsmath}

\usepackage{graphicx}
\usepackage[colorinlistoftodos]{todonotes}
\usepackage[sort]{natbib}
\usepackage[colorlinks=true, allcolors=blue]{hyperref}
\usepackage{authblk}
\usepackage{booktabs}
\usepackage{flafter}
\usepackage{lscape}
\usepackage{afterpage}
\usepackage{placeins}
\usepackage{wrapfig} 
\usepackage{algorithm}
\usepackage[noend]{algpseudocode}

\FloatBarrier
\title{The MetaSUB Microbiome Core Analysis Pipeline Enables Large Scale Metagenomic Analysis}

\author[1,2]{David C. Danko}
\author[2,3,$\ast$]{Christopher Mason}

\affil[1]{\small 
Tri-Institutional Computational Biology \& Medicine Program, Cornell University, NY, USA}
\affil[2]{\small Institute for Computational Biomedicine, Department of Physiology and Biophysics, Weill Cornell Medicine of Cornell University, NY, USA}
\affil[3]{\small The Feil Family Brain and Mind Research Institute, Weill Cornell Medicine, NY, USA}

\affil[$\ast$]{\small Corresponding author }

\date{}
\begin{document}
\maketitle

\begin{abstract}

\textbf{Motivation}  Accurate data analysis and quality control is critical for metagenomic studies. Though many tools exist to analyze metagenomic data there is no consistent framework to integrate and run these tools across projects. Currently, computational analysis of metagenomes is time consuming, often misses potentially interesting results, and is difficult to reproduce. Further, comparison between metagenomic studies is hampered by inconsistencies in tools and databases.

\textbf{Results} We present the MetaSUB Core Analysis Pipeline (CAP) a comprehensive tool to analyze metagenomes and summarize the results of a project. The CAP is designed in a bottom up fashion to perform QC, preprocessing, analysis and even to build relevant databases and install necessary tools.

\textbf{Availability and Implementation} The CAP is available under an MIT License on GitHub at \url{https://github.com/MetaSUB/CAP2} and on the Python Package Index. Documentation and examples are available on GitHub.

\end{abstract}

\section{Introduction}

Advances in metagenomics have led to a sharp increase in the number and size of studies. A large number of tools have been built to analyze metagenomes (in particular to generate taxonomic profiles) but there is still wide variation in how these tools are applied. This has created unnecessary friction as metagenomic studies often develop their pipelines in an ad-hoc fashion and as different choices for implementation (particularly choice of reference database) complicate cross study comparisons.

The CAP addresses issues of unnecessary friction in metagenomic analysis by providing a standardized analysis pipeline and framework. This provides researches with an effective starting point from which to launch more custom analyses. The CAP provides consistent naming and file structure conventions, clear documentation for different pipeline stages, automatic installation of constituent programs, automatic construction of reference databases and indices, project level summaries of results, comprehensive versioning system for reproducibility, and an easy to use python interface and API.

Critically the CAP does not attempt to replace existing analysis tools. The CAP is, rather, a set of standardized practices for metagenomic analysis based largely on existing peer reviewed tools. The flexible nature of the CAP also means that tools and databases can be swapped or modified with minimal effort.

\section{Materials and Methods}

\subsection{Installing and Running the CAP}

The CAP is written in Python and can be installed from the python package index using tools like \textit{pip} or \textit{easyinstall}. Once installed the CAP can be run from the command line or called from other programs via a python API. The CAP does not require users to install any constituent programs or databases. All necessary sub-programs are installed through PyPi or Conda and databases can be downloaded or built from source based on configuration. 

To run the CAP a user need only supply a CSV file with sample names and the locations of the forward and reverse FASTQ read file for each sample. Optionally the user can supply some configuration parameters (either through command line arguments or environment variables) which will direct the CAP to install programs and databases in specific locations. If for some reason installation of subprograms through conda is not available users can supply custom executables via configuration.

\subsection{Pipeline Stages and Tools}

\begin{figure}
  \begin{center}
    \includegraphics[width=0.8\textwidth]{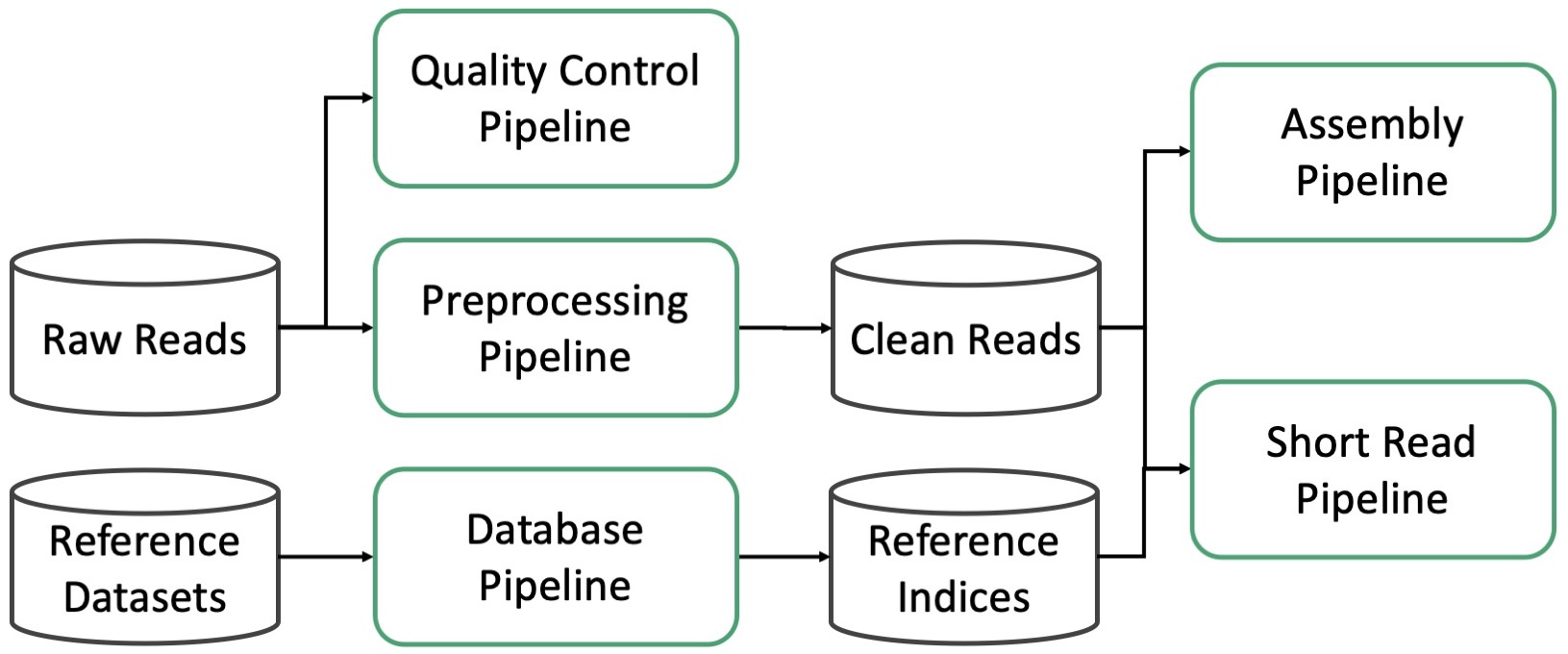}
	\caption{\small{
	    Overall structure of the pipeline. 
	}}
    \label{fig:pipe}
  \end{center}
\end{figure}

The CAP consists of 5 stages (Fig \ref{fig:pipe}): a quality control stage (Fig \ref{fig:qc}), a preprocessing stage (Fig \ref{fig:pre}), a short read stage (Fig \ref{fig:sr}), an assembly stage (Fig \ref{fig:assembly}), and a database stage (Fig \ref{fig:db}). The database stage consists of building relevant indices for the other stages. Since most users will not want to build databases the CAP is configured to download versioned databases by default.

The quality control stage consists of three modules: a custom module to count reads and associated statistics (e.g. read length, GC content, etc), FastQC \citep{Andrews2010}, and MultiQC \citep{Ewels2016}.

The preprocessing stage consists of 3 modules in series. First, adapters are removed from raw reads using AdapterRemoval \citep{Schubert2016}. Second, reads mapping to the human genome (hg38, with alternate contigs) are removed using Bowtie2 \citep{LangmeadandStevenLSalzberg2013}. Third, reads are error corrected using BayesHammer \citep{Nikolenko2013}.

The short read stage follows the preprocessing stage. It consists of several modules (or small groups of modules) that all start directly from pre-processed reads. Taxonomic profiling is performed by Kraken2 \citep{Wood2019} using a database based on all of RefSeq microbial followed by Bracken \citep{Lu2017}. Functional profiling is done with HUMANn2 \citep{Franzosa2018} after alignment of reads to UniRef90 using Diamond \citep{Buchfink2014}. MicrobeCensus \citep{Nayfach2015} is used to estimate the average genome size of a microbiome. MASH is used to generate small sketches of the data which are then used in a custom module to find the similarity to several human microbiomes \citep{Consortium2012}. A custom module is used to generate read statistics.

The assembly stage also follows the preprocessing stage. It consists of assembly using MetaSPAdes (omitting error correction since that is performed in the preprocessing stage) \citep{Nurk2017} followed by contig clustering using MetaBAT2 \citep{Kang2019}. ORFs are predicted in contig clusters using prodigal \citep{Hyatt2010}. Antimicrobial Resistance Genes are identified on contigs using RGI \citep{Alcock2020}.

After processing the CAP includes tools to build summary reports for a projects. Generally these reports consist of CSV files summarizing the outputs of tools across all samples in the project.

\begin{figure}
  \begin{center}
    \includegraphics[width=0.65\textwidth]{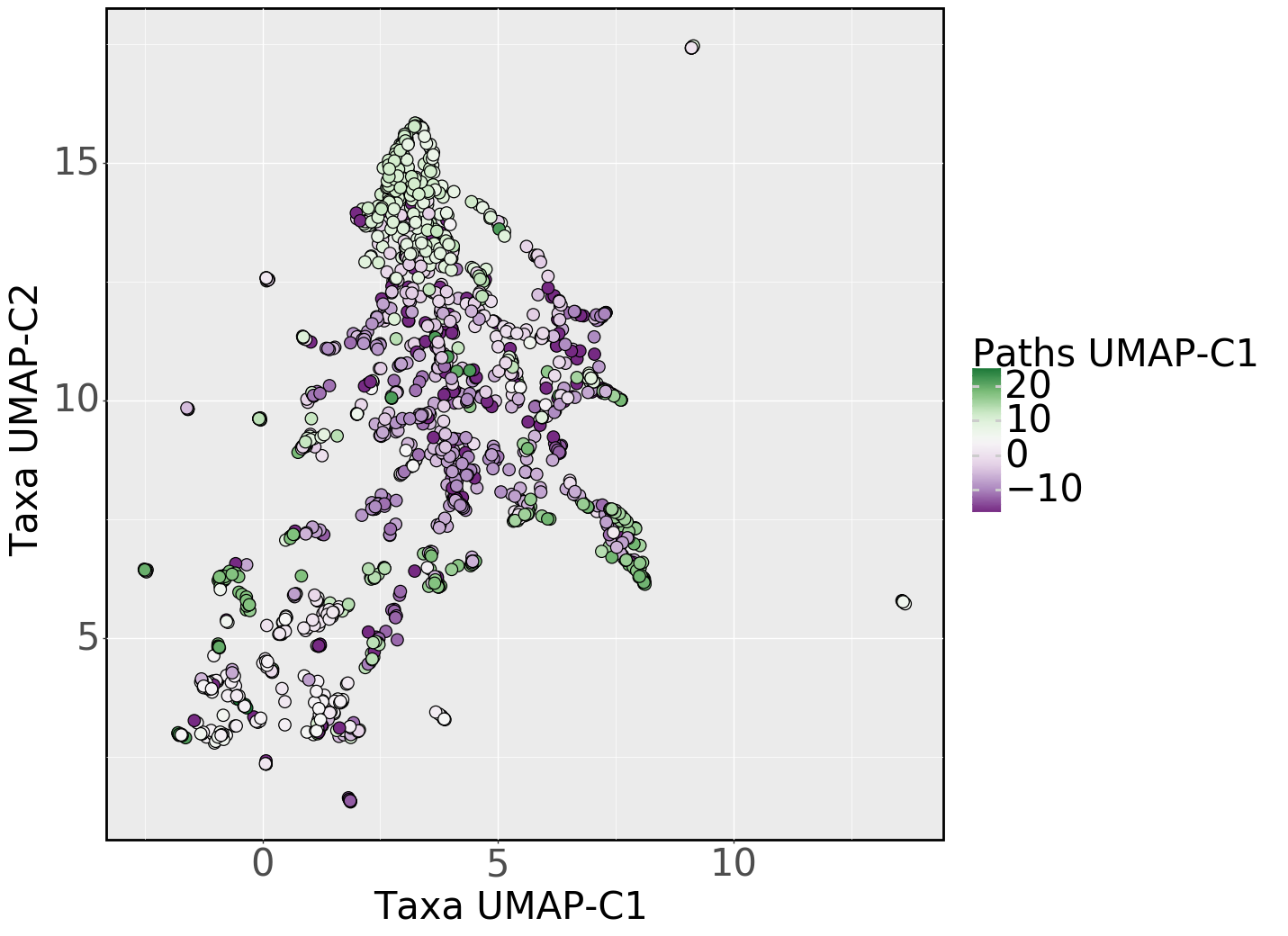}
	\caption{\small{
	    Relationship of taxonomic profiles and functional pathways. Position of each point shows the similarity of each samples taxonomic profile to other samples (nearby samples are more similar, reduced to 2-d with UMAP). Color of each point indicates similarity of functional profiles of samples (reduced to 1-d with UMAP). A rough relationship between taxonomic and function profiles can be seen.
	}}
    \label{fig:taxa}
  \end{center}
\end{figure}

\begin{figure}
  \begin{center}
    \includegraphics[width=0.5\textwidth]{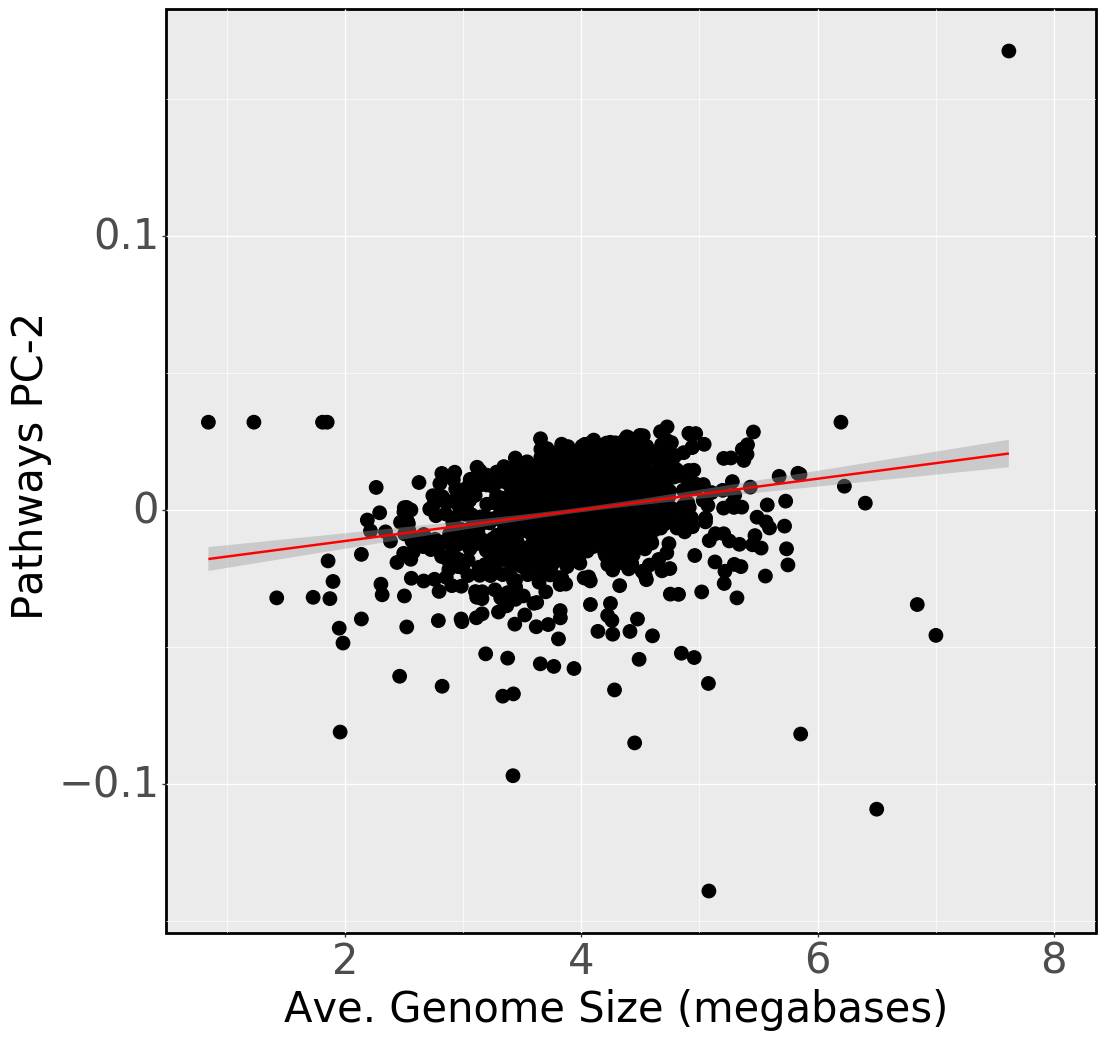}
	\caption{\small{
	    Relationship between functional pathways and estimated average genome size. The estimated average genome size is positively correlated (rho 0.24, p < 2e-16) to the second principal component of the functional profile.
	}}
    \label{fig:ags}
  \end{center}
\end{figure}

\subsection{Modules, Versioning, and Reproducibility}

The CAP is composed of modules each of which corresponds (roughly) to a single sub-programs. These modules contain documentation for the specific sub-program, specify dependencies, outputs and how the sub-program will be run. Modules are python classes and may be dynamically modified e.g. to substitute different module versions or to insert stages for experimentation. This makes it easy to develop and add modules to the CAP.

A major feature of the CAP is its strict multi-layered versioning system. Each module specifies a specific version number (typically a SemVer) and dependencies. To account for upstream changes modules also specify a version tree that includes the version numbers of all upstream modules (and their upstream modules, etc) into a tree. This tree can be accessed either as a human readable newick tree or as a version hash. When run each modules records the version number of the module run, the hash of the version tree, and the version of the sub-program in the module (in case a novel version has been specified in the config) as a JSON file alongside the actual module results. This system enables comparison and reproducibility between pipeline versions.

\section{Example Analysis}

To demonstrate some of the capabilities of the CAP we analyzed 1,521 environmental metagenomic samples from the PathoMAP project \citep{Afshinnekoo2015}. These samples were collected from dry surfaces in the New York City Subway system in 2015.

To demonstrate the CAP we present two results that show how multiple tools in the CAP can be combined to produce novel results. In particular we show two results from three different tools that are part of the CAP: functional profiling using HUMANn2, taxonomic profiling with Kraken2, and average genome size estimation using MicrobeCensus. Our goal with these results is to show how multiple tools can be used to uncover certain ecological relationships in metagenomic data.

We reduced the dimensionality of our taxonomic profiles to 2 using UMAP and analagously reduced functional profiles to 1 dimension. The functional profiles and taxonomic profiles show some relation (Figure \ref{fig:taxa}) with samples that have similar taxonomic profiles also having similar functional profiles. Average genome size of bacteria in microbiomes has been postulated to have some connection to metabolic capacity with larger genomes in more generalist bacteria \citep{Nayfach2015}. We show a relationship between average genome size and functional capacity (Figure \ref{fig:ags}) with a weak correlation (spearmans rho 0.24, p < 2e-16) between average genome size and the second principal component of the functional profile.

\section{Conclusion}

The CAP provides a consistent good-practice baseline for Metagenomic studies. It is based, largely off of existing peer reviewed tools that have themselves been extensively benchmarked. Routine use of the CAP is meant to improve reproducibility in metagenomic studies, decrease friction in performing such studies, and enable cross-project comparisons. The routine use of multiple analysis tools can result in novel insights and help to uncover ecological relationships.

\section{Acknowledgement}

We would like to thank Robert Petit and Ben Chrobot for their help.

\setcitestyle{semicolon}
\bibliographystyle{apalike}
\bibliography{bib}

\newpage

\section*{Supplement}

\setcounter{figure}{0}
\makeatletter 
\renewcommand{\thefigure}{S\@arabic\c@figure}
\makeatother

\begin{figure}
  \begin{center}
    \includegraphics[width=0.4\textwidth]{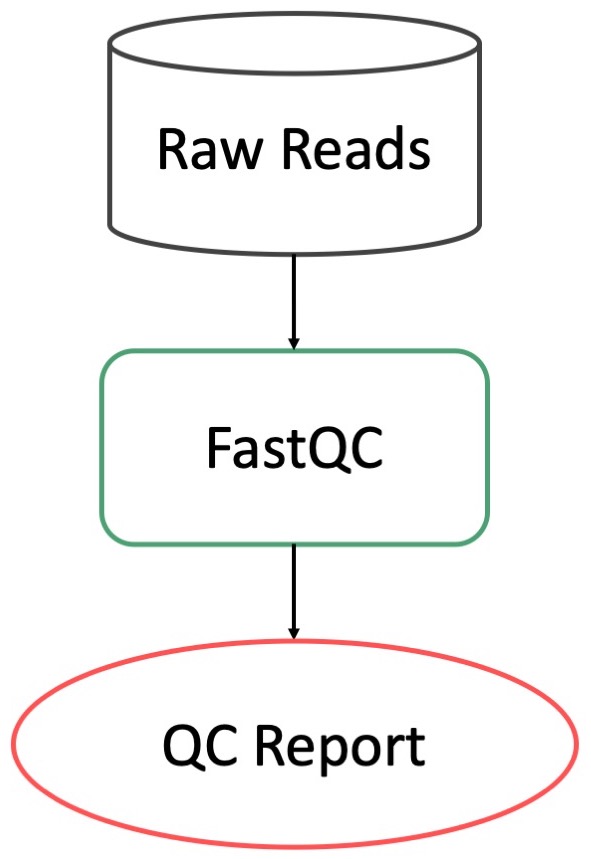}
	\caption{\small{
	    Quality control stage, black cylinders represent raw data, orange ovals are results, and green boxes are tools.
	}}
    \label{fig:qc}
  \end{center}
\end{figure}

\begin{figure}
  \begin{center}
    \includegraphics[width=0.8\textwidth]{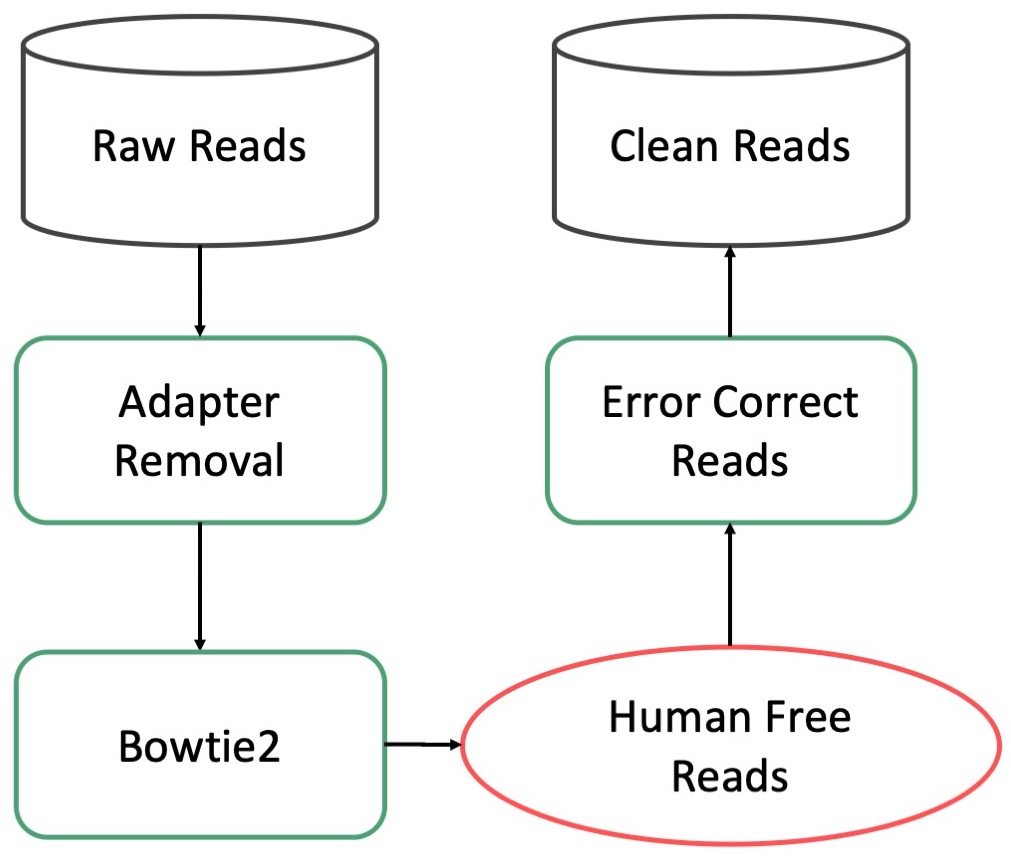}
	\caption{\small{
	    Preprocessing stage, black cylinders represent raw data, orange ovals are results, and green boxes are tools.
	}}
    \label{fig:pre}
  \end{center}
\end{figure}

\begin{figure}
  \begin{center}
    \includegraphics[width=0.95\textwidth]{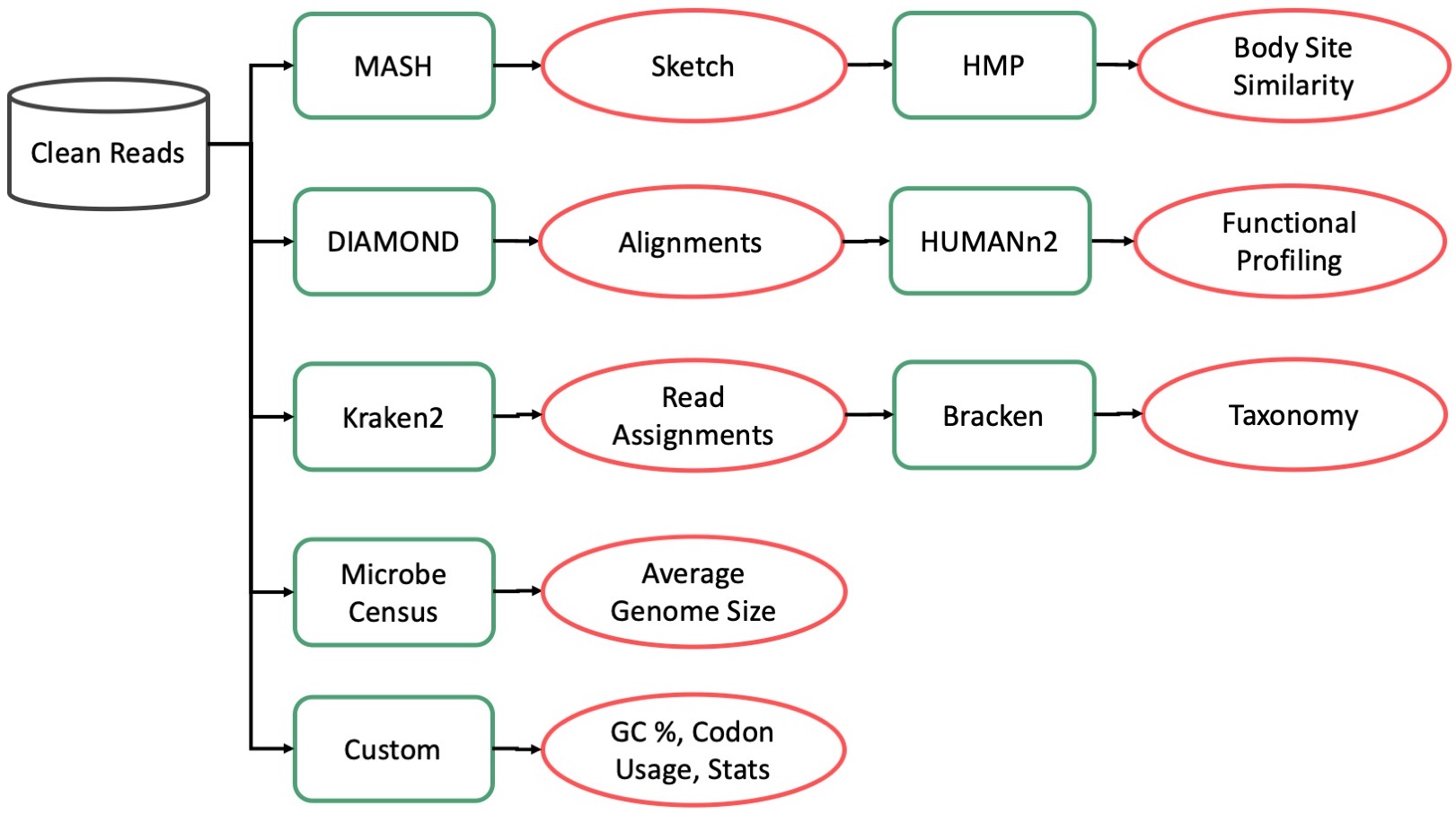}
	\caption{\small{
	    Short Read stage, black cylinders represent raw data, orange ovals are results, and green boxes are tools.
	}}
    \label{fig:sr}
  \end{center}
\end{figure}

\begin{figure}
  \begin{center}
    \includegraphics[width=0.95\textwidth]{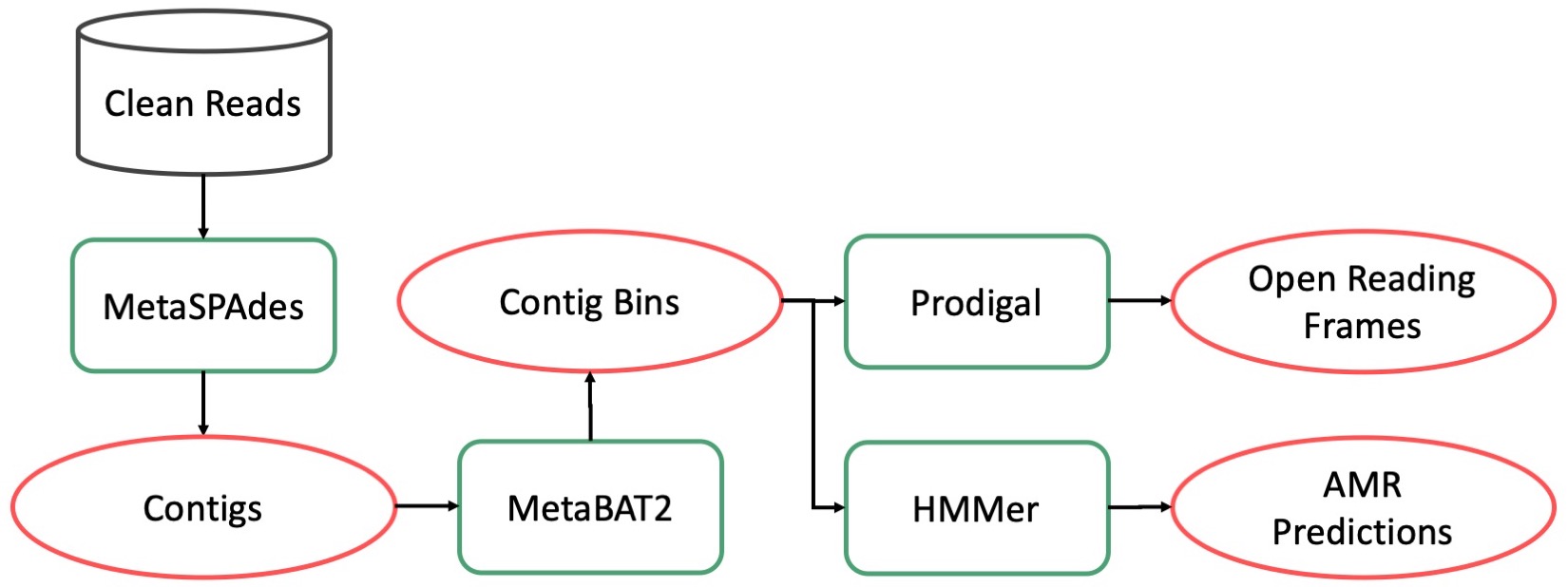}
	\caption{\small{
	    Assembly stage, black cylinders represent raw data, orange ovals are results, and green boxes are tools.
	}}
    \label{fig:assembly}
  \end{center}
\end{figure}

\begin{figure}
  \begin{center}
    \includegraphics[width=0.8\textwidth]{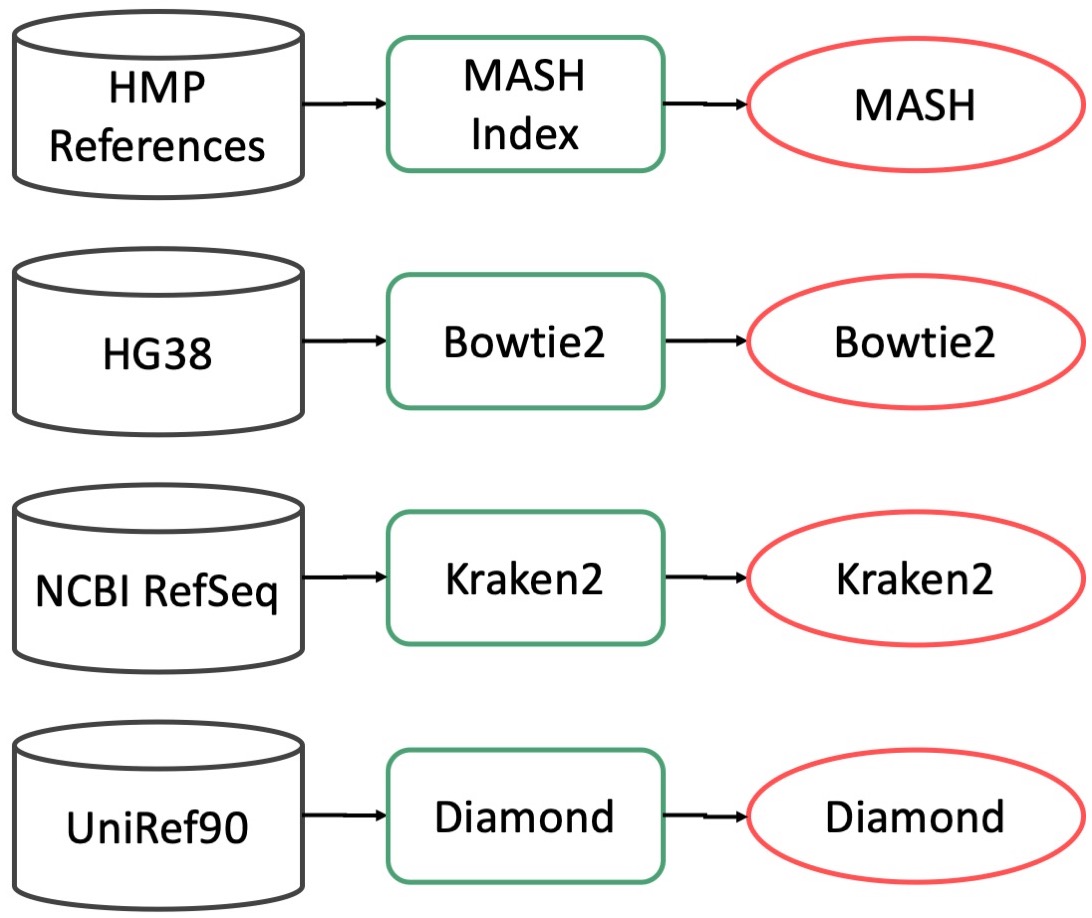}
	\caption{\small{
	    Database stage, black cylinders represent raw data, orange ovals are results, and green boxes are tools.
	}}
    \label{fig:db}
  \end{center}
\end{figure}

\end{document}